\documentclass{sig-alternate-05-2015}

\usepackage{subfigure}
\usepackage[hidelinks]{hyperref}

\graphicspath{{figures/}}
\DeclareGraphicsExtensions{.pdf,.jpeg,.png}

\makeatletter
\def\@copyrightspace{\relax}
\makeatother

\begin{document}

\title{Impact of Three-Dimensional Video Scalability on Multi-View Activity Recognition using Deep Learning\titlenote{\textsf{Accepted as a Thematic Workshops paper at ACM MM 2017.}}}

\numberofauthors{2}

\author{
	\alignauthor Jun-Ho Choi \\[0.3em]
	\affaddr{School of Integrated Technology} \\
	\affaddr{Yonsei University} \\
	\affaddr{Incheon, South Korea} \\[0.3em]
	\email{idearibosome@yonsei.ac.kr}	
	\alignauthor Manri Cheon \\[0.3em]
	\affaddr{School of Integrated Technology} \\
	\affaddr{Yonsei University} \\
	\affaddr{Incheon, South Korea} \\[0.3em]
	\email{manri.cheon@yonsei.ac.kr}
	\and	
	\alignauthor Min-Su Choi \\[0.3em]
	\affaddr{School of Integrated Technology} \\
	\affaddr{Yonsei University} \\
	\affaddr{Incheon, South Korea} \\[0.3em]
	\email{min-su.choi@yonsei.ac.kr}	
	\alignauthor Jong-Seok Lee \\[0.3em]
	\affaddr{School of Integrated Technology} \\
	\affaddr{Yonsei University} \\
	\affaddr{Incheon, South Korea} \\[0.3em]
	\email{jong-seok.lee@yonsei.ac.kr}
}

\maketitle

\begin{abstract}
	
Human activity recognition is one of the important research topics in computer vision and video understanding.
It is often assumed that high quality video sequences are available for recognition.
However, relaxing such a requirement and implementing robust recognition using videos having reduced data rates can achieve efficiency in storing and transmitting video data.
Three-dimensional video scalability, which refers to the possibility of reducing spatial, temporal, and quality resolutions of videos, is an effective way for flexible representation and management of video data.
In this paper, we investigate the impact of the video scalability on multi-view activity recognition.
We employ both a spatiotemporal feature extraction-based method and a deep learning-based method using convolutional and recurrent neural networks.
The recognition performance of the two methods is examined, along with in-depth analysis regarding how their performance vary with respect to various scalability combinations.
In particular, we demonstrate that the deep learning-based method can achieve significantly improved robustness in comparison to the feature-based method.
Furthermore, we investigate optimal scalability combinations with respect to bitrate in order to provide useful guidelines for an optimal operation policy in resource-constrained activity recognition systems.

\end{abstract}


\keywords{Activity recognition, video scalability, deep learning}

\maketitle

\section{Introduction}

Automated human activity recognition tasks using videos have been extensively studied in the last decades, since they can be applied to a wide range of applications such as surveillance \cite{vishwakarma2013survey}, smart home \cite{van2008accurate}, and health care \cite{avci2010activity}.
However, it is still challenging to achieve good recognition performance due to the highly complex nature of the tasks and diverse conditions of the recording environment.
To tackle these, many activity recognition methods have been proposed in literature \cite{herath2017going}.
While most algorithms have focused on finding distinct characteristics of activities from videos based on several feature descriptors, deep learning-based approaches have been developed recently and shown promising results \cite{herath2017going}.

Many activity recognition algorithms have been studied in "lab settings", i.e., highly controlled environments with abundant computing and network resources.
However, when it comes to the real world, the efficiency of a recognition system is an important issue for ensuring, e.g., real-time low-delay operation, low-cost implementation, and low-power operation.
In particular, since the overheads required to store and transmit high quality video data are usually significant, ways to reduce the video data rate are frequently involved in real-world applications, e.g., compression and downscaling.
Unfortunately, reduction of the video data rate tends to cause degradation of video quality, which in turn may lower the recognition performance.
This is particularly problematic when the recognizer has been trained using high quality videos but is deployed to a system where only lower quality videos are available due to resource constraints.
In this context, it is essential to study how the recognition performance is influenced by video data reduction.
In general, reduction of the video data rate can be achieved by reducing spatial resolution (i.e., frame size), temporal resolution (i.e., frame rate), and quality resolution (i.e., quantization during compression), which is called three-dimensional video scalability \cite{lee2012quality}.

There have been some studies examining the influence of video scalability on feature extraction-based activity recognition methods \cite{harjanto2016investigating,see2015effects,srinivasan2016robustness}.
However, in-depth analysis of deep learning-based activity recognition with respect to multi-dimensional video scalability has not been reported.
In this paper, we investigate the impact of the multi-dimensional video scalability, including quantization parameter, spatial resolution, and frame rate, on both a conventional feature extraction-based method and a state-of-the-art deep learning method for activity recognition.
In particular, we aim to provide insight regarding the optimal scalability combinations when both the data rate and recognition performance are considered.
We employ a dataset containing multi-view videos, which is beneficial to additionally explore the content dependency, and analyze the performance changes for diverse scalability combinations.

The rest of the paper is organized as follows.
We review related works in Section 2.
Section 3 describes the dataset, models, and scalability combinations that we utilize.
We present our in-depth analysis of the results in Section 4 and conclude the paper in Section 5.

\section{Related Work}

Automated activity recognition has been studied in many different application contexts \cite{herath2017going}.
Traditionally, a dominant approach is based on extracting representative features from video frames.
Laptev \textit{et al.} introduced a bag-of-features (BoF) approach by extracting space-time interest points (STIPs) and aggregating feature descriptors by computing histograms of the visual words \cite{laptev2008learning}. 
Wang \textit{et al.} suggested descriptors based on dense trajectories and motion boundary histograms \cite{wang2013dense}.
Recently, deep learning architectures have begun to be employed and been shown to yield improved recognition performance.
Ji \textit{et al.} extended the convolutional neural networks to three-dimension to deal with both spatial and temporal characteristics of activities in videos \cite{ji20133d}.
Donahue \textit{et al.} built a deep learning model having both convolutional and recurrent neural networks to capture the spatial and temporal characteristics, respectively \cite{donahue2015long}.

In general, videos can have various combinations of spatial, temporal, and quality resolutions, which is noted as video scalability.
Previous studies in several fields investigated how it affects the performance of the video systems, including surveillance \cite{klare2010assessment,mcguinness2015beyond}, face recognition \cite{boom2006effect,klare2010assessment}, visual perception of users \cite{zhai2008cross,lee2012quality,cheon2016evaluation}, and so on.
The video scalability has been also considered in human activity recognition tasks.
Harjanto \textit{et al.} investigated the impact of different frame rates on human activity recognition in several feature extraction-based approaches, and observed degradation of recognition performance for reduced frame rates \cite{harjanto2016investigating}.
See and Rahman analyzed the impact of video scalability on activity recognition using extremely low quality videos \cite{see2015effects}.
Srinivasan \textit{et al.} analyzed the performance of a method using compression-domain features (i.e., motion vectors), and confirmed the fluctuation of performance with respect to spatial and quality variations \cite{srinivasan2016robustness}.
However, these studies focused only on feature extraction-based methods and furthermore, did not consider various combinations of the video scalability dimensions thoroughly.

\section{Method}

\subsection{Dataset}

\begin{figure}[t]
	\centering
	\subfigure[]{
		\includegraphics[width=1.5in]{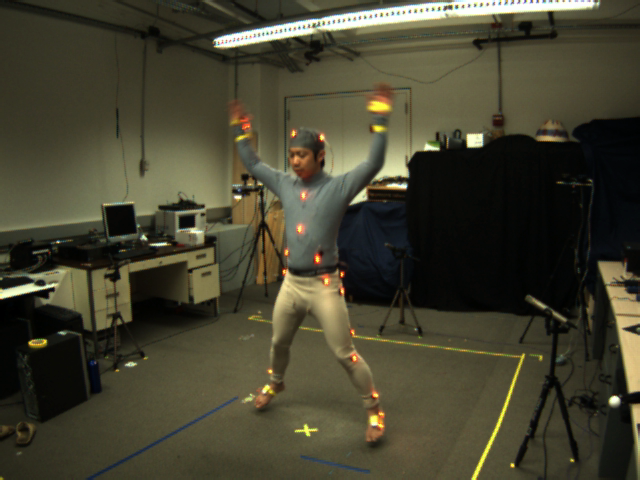}
	}
	\subfigure[]{
		\includegraphics[width=1.5in]{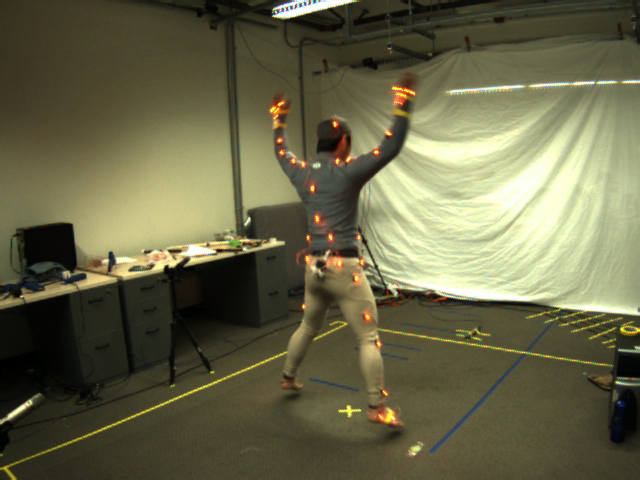}
	}
	\subfigure[]{
		\includegraphics[width=1.5in]{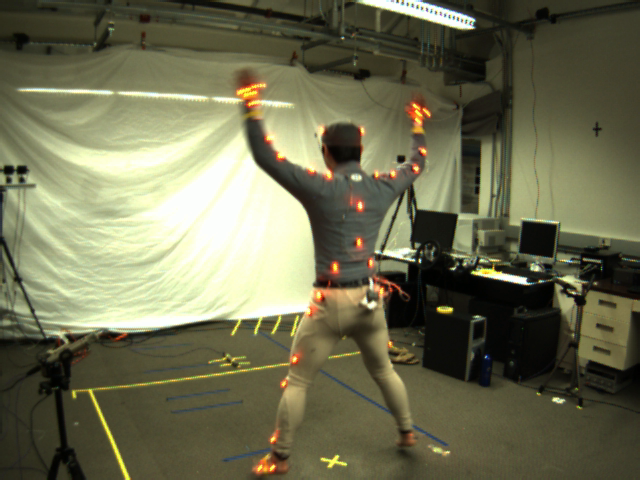}
	}
	\subfigure[]{
		\includegraphics[width=1.5in]{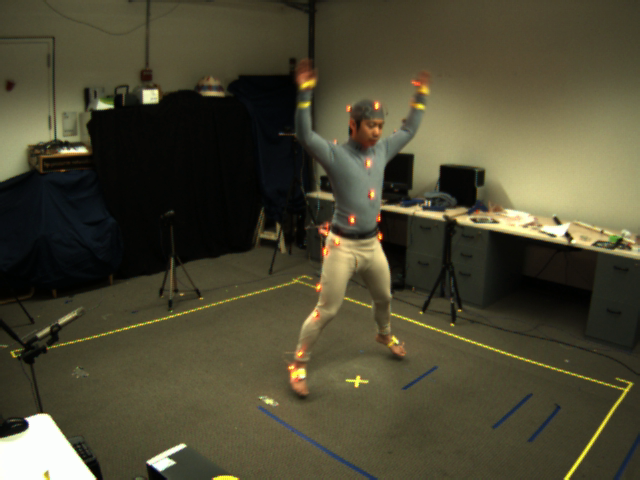}
	}
	\caption{Examples frames of the Berkeley MHAD dataset \cite{ofli2013berkeley}. (a) Cluster 1 (b) Cluster 2 (c) Cluster 3 (d) Cluster 4}
	\label{fig:example_frames}
\end{figure}

Our investigation focuses on multi-view activity recognition.
To achieve this, we employed the Berkeley Multimodal Human Action Database (Berkeley MHAD) \cite{ofli2013berkeley}, which has multiple videos simultaneously recorded from 12 cameras.
It consists of 11 activities, including \textit{jumping}, \textit{jumping jacks}, \textit{bending}, \textit{punching}, \textit{waving hands}, \textit{waving one hand}, \textit{clapping}, \textit{throwing a ball}, \textit{sit down then stand up}, \textit{sit down}, and \textit{stand up}.
The activities were performed by 12 subjects and repeated five times for each activity.
The cameras were arranged into four different clusters.
Four cameras were placed in clusters 1 and 4, and two cameras were placed in clusters 2 and 3.
Figure~\ref{fig:example_frames} shows an example frame for each cluster.
Since the dataset contains videos recorded in different points of view, it is possible to examine viewpoint-dependent impacts of video scalability.
In this work, the videos of the first seven subjects are used for training the recognizers, and those of the last five subjects are used for evaluating the performance of the trained recognizers, as suggested in \cite{ofli2013berkeley}.

\subsection{Models}

\begin{figure*}[t]
	\centering
	\includegraphics[width=6.9in]{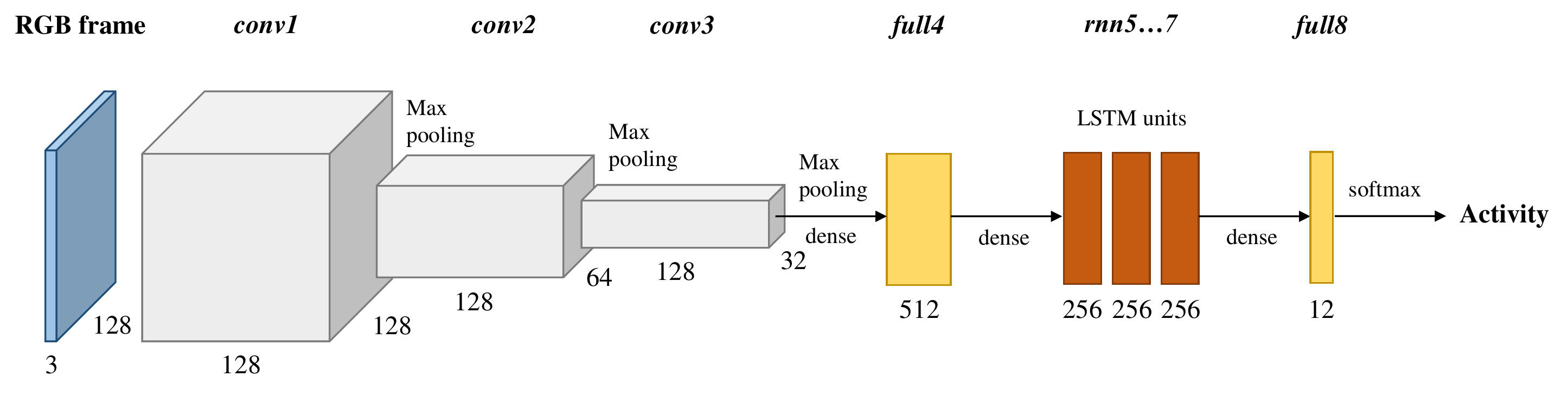}
	\caption{Our deep learning model for activity recognition.}
	\label{fig:deep_learning_model}
\end{figure*}

\subsubsection{Feature extraction-based model}

We consider a feature extraction-based activity recognition method based on the histogram of gradients (HoG) and histogram of flow (HoF) extracted by the STIP detector \cite{laptev2008learning}, which was also used in \cite{ofli2013berkeley}.
For each video, HoG and HoF features are extracted.
Then, the extracted features for the training dataset are aggregated to form 20 groups by the $k$-medoids algorithm, where the Euclidean distance is used as the distance measure.
The representative feature, i.e., the medoid, in each group is regarded as the "codeword" of the group.
A given feature is assigned to the codeword that has the minimum Euclidean distance to the feature.
Then, the video is represented as a histogram of codewords to which the features in the video are assigned.
From the histograms of codewords, a kernel-SVM (K-SVM) is constructed for each cluster with the $\chi^2$ kernel given by
\begin{equation}
k(m, n) = 1 - \frac{1}{2} \sum_{i} \frac{({m}_{i}-{n}_{i})^2}{({m}_{i}+{n}_{i})}
\end{equation} 
where ${m}_{i}$ and ${n}_{i}$ are the $i$-th bins of two histograms $m$ and $n$, respectively.
In our implementation, the LIBSVM library \cite{chang2011libsvm} was used.
For multi-class classification using binary K-SVMs, a one-versus-one scheme implemented in the library was used.

\subsubsection{Deep learning-based model}

We built an end-to-end deep learning model that contains both convolutional and recurrent layers.
Figure~\ref{fig:deep_learning_model} shows an overview of our deep learning model.
The model consists of three convolutional layers ($conv1$, $conv2$, $conv3$), one dense layer ($full4$), three recurrent dense layers ($rnn5$, $rnn6$, $rnn7$), and one softmax dense layer ($full8$).
For the convolution layers, the rectified linear unit (ReLU) function (i.e., $R(x) = \max(x, 0)$) is used as the activation function, and the resolution of each output is reduced by a factor of two via max pooling.
As the units in the recurrent layers, long short-term memory (LSTM) \cite{hochreiter1997long} cells are used.
Note that the number of outputs at the final layer is 12, which is one more than the number of activities.
The first output corresponds to \textit{no action}, and the rest correspond to each of the 11 activities.
We implemented the model in TensorFlow 1.0\footnote{https://www.tensorflow.org}.

Our model is similar to the long-term recurrent convolutional network (LRCN) in \cite{donahue2015long} and DeepConvLSTM in \cite{ordonez2016deep}.
In detail, the convolutional layers operate as \textit{feature extractors} -- they find several features that are effective to distinguish different actions, and the recurrent layers operate as \textit{action trackers} -- they track the change of the features across adjacent frames and keep useful information to classify the activity for each frame.
Thanks to this distinguished structure, it is possible to inspect what the roles of the convolutional and recurrent layers are to recognize activities cooperatively.

In the training stage for a camera cluster, a video is randomly chosen and at most 110 consecutive frames (i.e., five seconds in length) are extracted.
Then, each frame is cropped with an arbitrary size within 400$\times$400 to 480$\times$480 pixels and resized to 128$\times$128 pixels.
After all frames are propagated through the layers of the model, the loss is calculated in terms of the cross-entropy function.
The Adam optimization method \cite{kingma2014adam} is used for updating the model parameters.
An individual model is constructed for each cluster.
We trained each model on GPUs having 1,280 cores and 6GB RAMs each.

Unlike the aforementioned K-SVM model, the deep learning model outputs a recognized activity class for every frame in the video.
It enables recognizing activity even in the middle of a video sequence, which is beneficial to build real-time recognition systems.
The final recognition result for a given video is obtained by aggregating the recognition results of all the frames as follows.
Let $p_{a,t,c}$ be the probability of activity $a$ at time $t$ for the video taken by $c$-th camera in the cluster, which is obtained from the softmax function after the \textit{full8} layer.
Then, the recognized activity $a_{t}^{*}$ at time $t$ is calculated as
\begin{equation}
a_{t}^{*} = \textrm{arg}\max_{a} \prod_{c}{p_{a,t,c}}.
\end{equation}
The final activity $a^{*}$ is determined by finding the most frequent activity through the entire video, i.e.,
\begin{equation}
a^{*} = \textrm{arg}\max_{a} \sum_{t}{\delta (a_{t}^{*}, a)}
\end{equation}
where $\delta (i, j) = 1$ if $i = j$ and $0$ if $i \neq j$.

\subsection{Scalability dimensions}

We consider various combinations of all the three dimensions of video scalability, i.e., quantization parameter (QP), spatial resolution, and frame rate.

\subsubsection{Quantization parameter}

In video coding methods, the quantization process is involved to reduce the bitrate of encoded video with sacrificing the visual quality.
The strength of quantization is controlled by the QP in modern video coding methods.

In H.264/AVC coding, the range of QP values is between 0 and 51, where larger QP values produce videos having lower data rates at the cost of more degraded quality \cite{wiegand2003overview}.
The QP value of zero produces lossless videos without quality degradation.
In our experiment, QP values of 0, 36, 41, 46, and 51 are used, and the period of intra frames is set to one second.

\begin{figure}[t]
	\centering
	\subfigure[]{
		\includegraphics[width=1.5in]{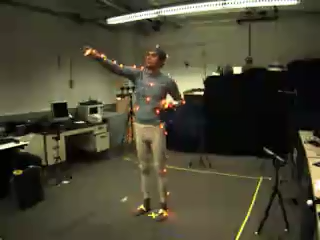}
	}
	\subfigure[]{
		\includegraphics[width=1.5in]{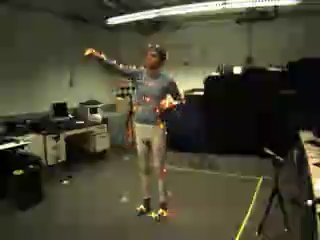}
	}
	\subfigure[]{
		\includegraphics[width=1.5in]{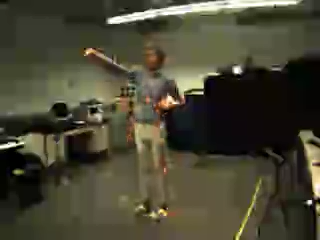}
	}
	\subfigure[]{
		\includegraphics[width=1.5in]{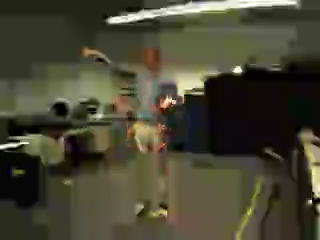}
	}
	\caption{Example frames of the encoded videos. (a) $QP=36$ (b) $QP=41$ (c) $QP=46$ (d) $QP=51$}
	\label{fig:example_qp_values}
\end{figure}

Figure~\ref{fig:example_qp_values} shows example frames encoded with various QP values.
As the QP value increases, the visual quality tends to decrease.
For example, it is noticeable that the face becomes unrecognizable, and the legs become indistinguishable from the floor due to their similar colors.
In addition, the blocking artifacts are prominent for $QP=51$, due to the segmentation of macroblocks in the encoding process \cite{wiegand2003overview}.

\subsubsection{Spatial resolution}
\label{sec:resolution}

We produced videos having resolutions of 320$\times$240, 160$\times$120, and 80$\times$60 pixels from the original videos having a resolution of 640$\times$480 pixels.
For the feature extraction-based model, reduced spatial resolutions can directly affect the recognition performance, because of the loss of describable features in each frame.
In the deep learning model, reducing the resolution to 320$\times$240 pixels does not affect the performance since the input size (128$\times$128 pixels) is still smaller, while the other resolutions may reduce the performance due to the interpolation.
However, when the quality degradation due to compression is additionally involved by adjusting the QP value, reducing the resolution to 320$\times$240 pixels may also influence the accuracy of the activity recognition.

\subsubsection{Frame rate}

The temporal scalability is realized by adjusting the frame rate.
For the original videos having a frame rate of 22 fps, we dropped one per two frames to produce videos at a frame rate of 11 fps and three per four frames to produce videos at a frame rate of 5.5 fps.
Since both the feature extraction-based and deep learning models depend on the temporal consistency of activities, the reduced frame rate may degrade the overall performance of the activity recognition.
In addition, reducing the frame rate can decrease the computational complexity for both activity recognition models, since the number of frames per video to be processed is reduced.

\section{Results}
\label{sec:results}

The evaluation of the two models was performed with videos having 60 scalability combinations (5 QPs $\times$ 4 frame sizes $\times$ 3 frame rates).
In total, we tested about 66,000 videos for each of camera clusters 1 and 4, and about 33,000 videos for each of camera clusters 2 and 3.
In the feature extraction-based model, the recognition was performed by extracting STIPs for a given video, calculating the histogram of the labeled codewords, and finding the matched activity using the trained K-SVM.
We also tested the nearest neighbor classifier as in \cite{ofli2013berkeley} but did not included in this paper, since it showed very similar performance to that of K-SVM.
In the deep learning model, each frame was cropped to keep only the central region and resized to a resolution of 128$\times$128 pixels to be served as the input.

\begin{figure*}[t]
	\centering
	\subfigure[]{
		\includegraphics[width=3.38in]{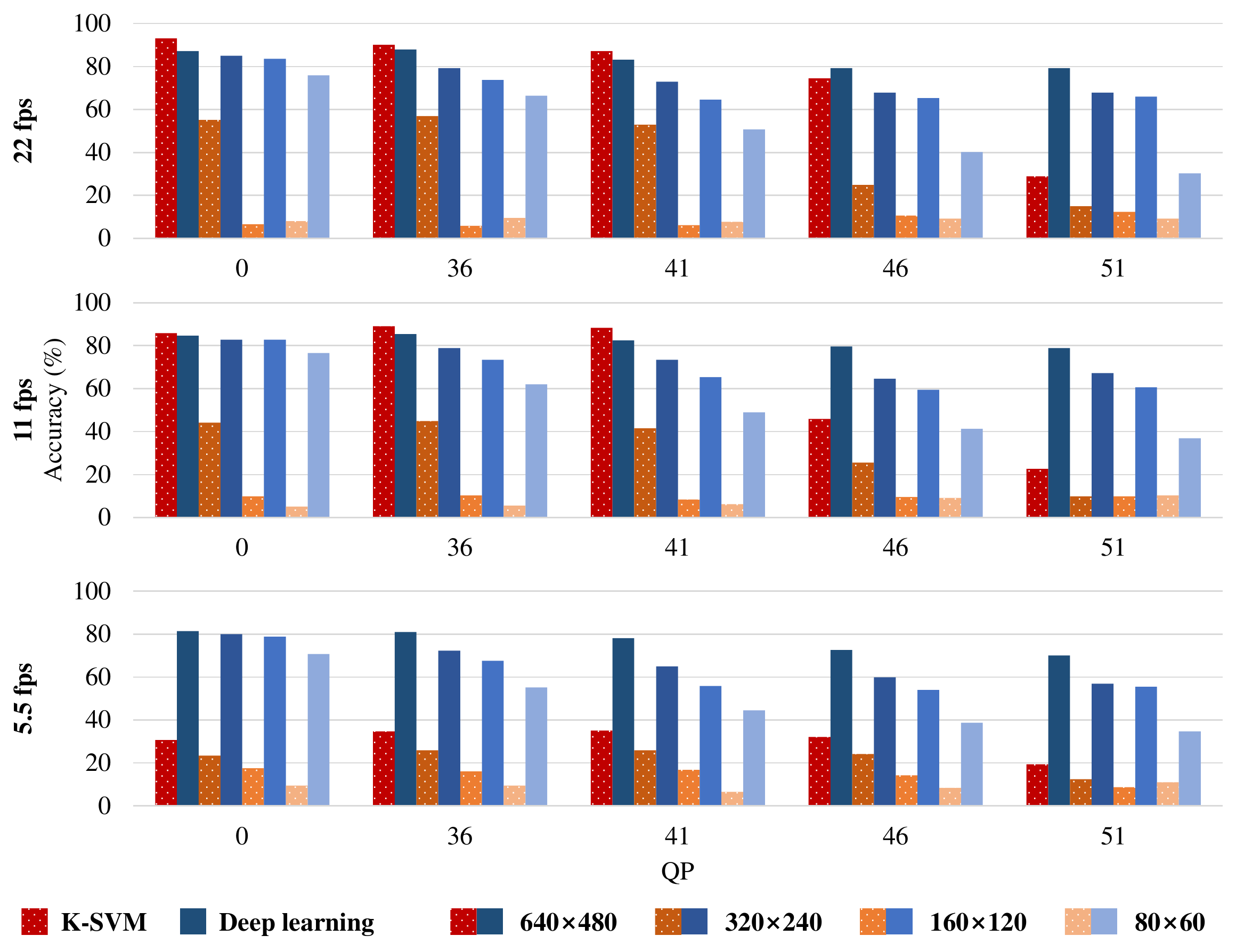}
	}
	\subfigure[]{
		\includegraphics[width=3.38in]{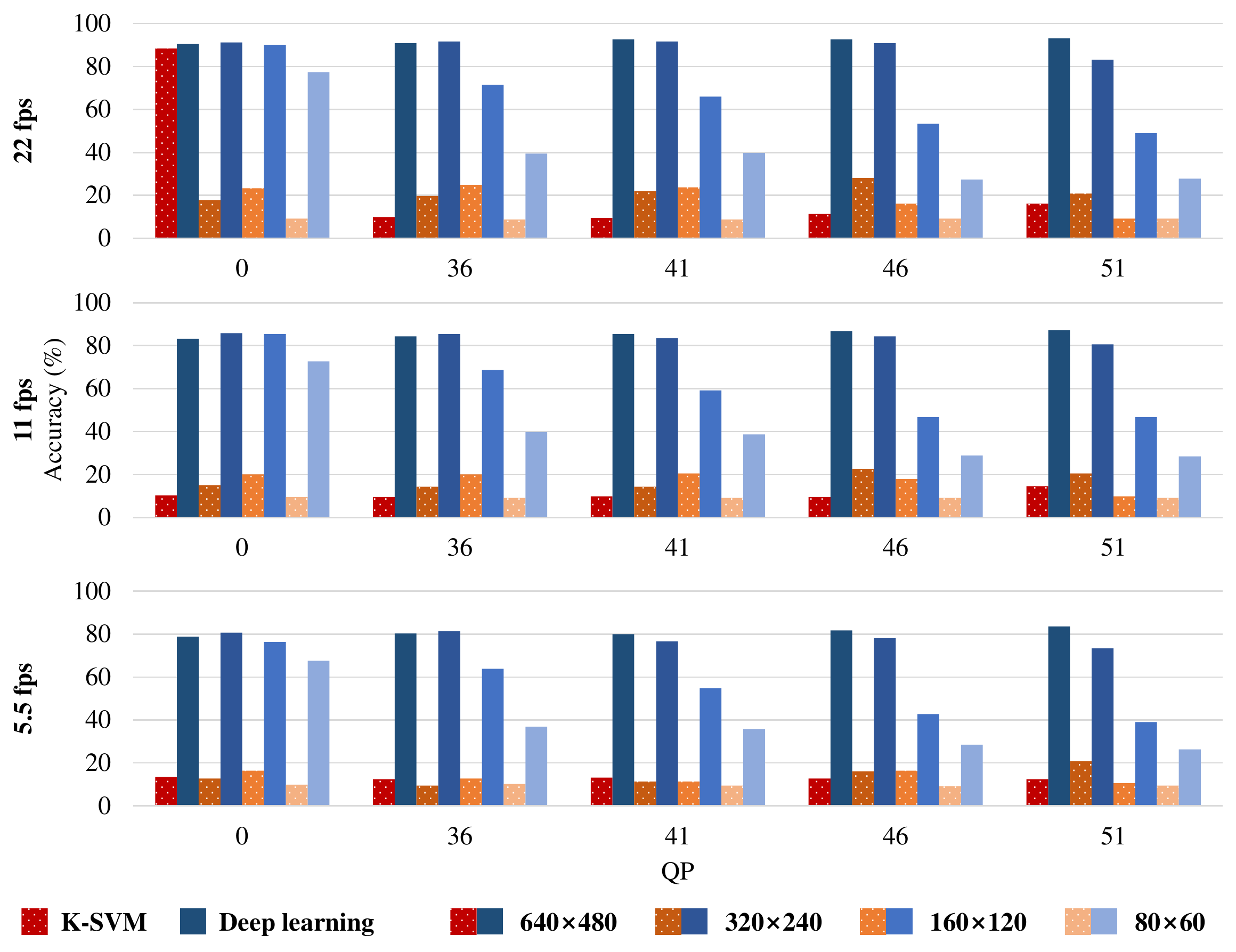}
	}
	\subfigure[]{
		\includegraphics[width=3.38in]{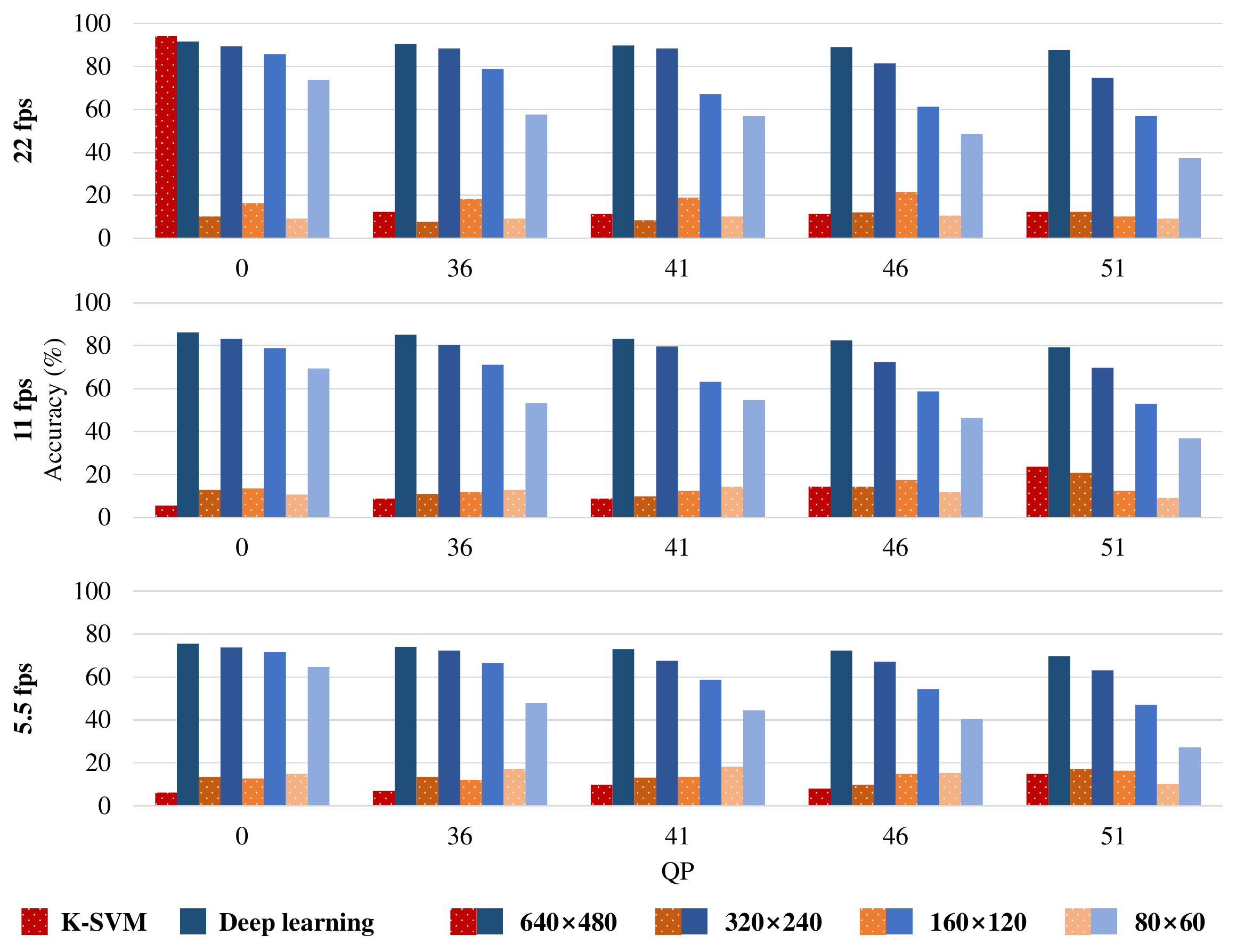}
	}
	\subfigure[]{
		\includegraphics[width=3.38in]{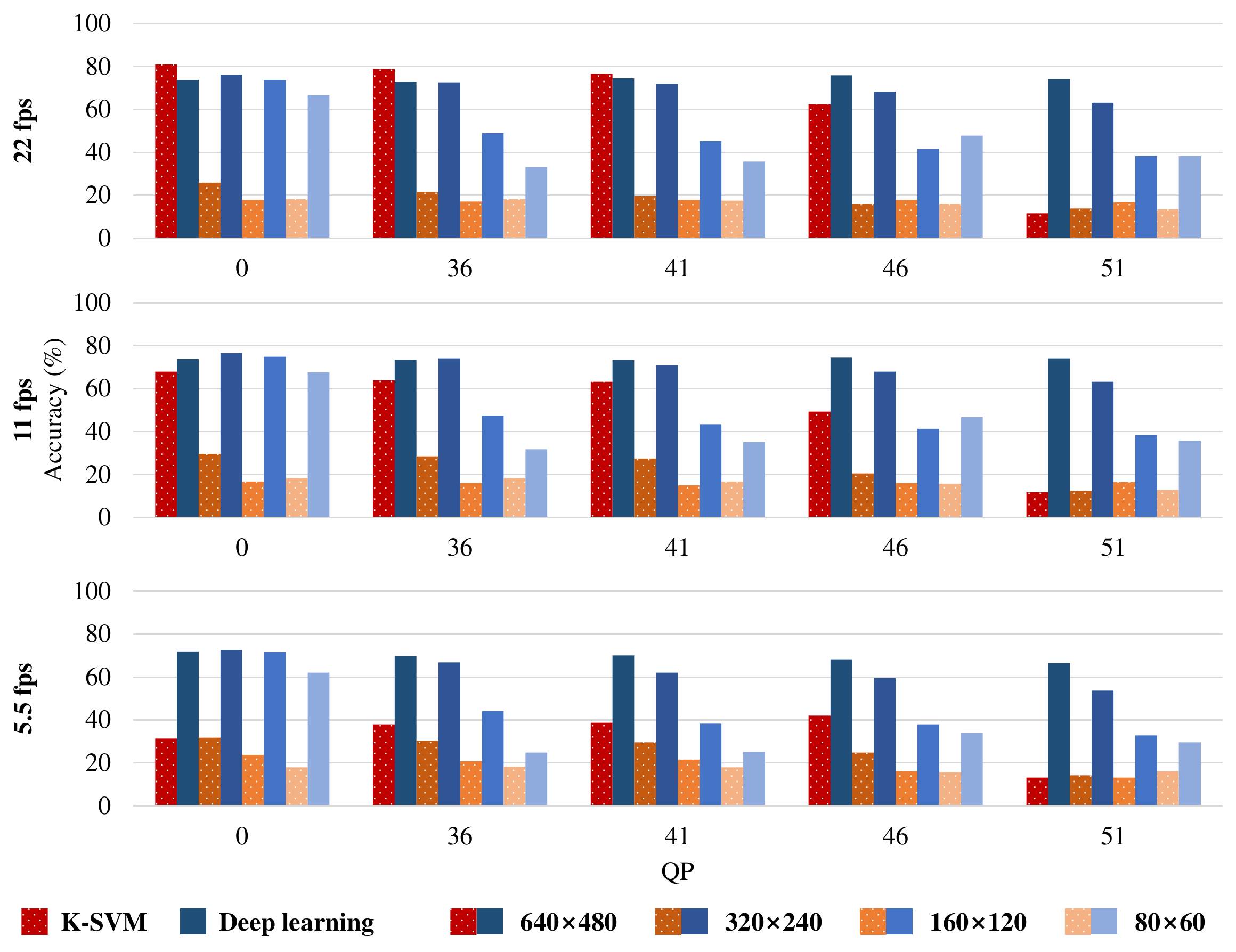}
	}
	\caption{Accuracy of the activity recognition. (a) Cluster 1 (b) Cluster 2 (c) Cluster 3 (d) Cluster 4}
	\label{fig:accuracy_per_clusters}
\end{figure*}

Figure~\ref{fig:accuracy_per_clusters} summarizes the accuracy of the activity recognition with respect to video scalability combinations for each cluster.
For the original videos (i.e., 22 fps, 640$\times$480 pixels, $QP=0$), the K-SVM and deep learning models achieve similar recognition performance.
Both models have the lowest performance at cluster 4, which is probably due to the farthest distance between the cameras and the subjects, as shown in Figure~\ref{fig:example_frames}.

For the degraded videos, however, the deep learning model outperforms the feature extraction-based model in most cases.
Moreover, in clusters 2 and 3, the accuracies of the K-SVM model for the degraded videos are significantly lower than those of the deep learning model.
For example, the highest accuracy achieved by the K-SVM model for the degraded videos in cluster 3 is 23.72\% when the spatial resolution is 640$\times$480 pixels, the frame rate is 11 fps, and the QP value is 51, which is even lower than the lowest accuracy obtained from the deep learning model for the degraded videos in cluster 3 (27.37\% when the spatial resolution is 80$\times$60 pixels, the frame rate is 5.5 fps, and the QP value is 51).
As shown in Figure~\ref{fig:example_frames}, the videos in clusters 2 and 3 were recorded at the back side of the subjects.
This demonstrates that depending on the camera viewpoint, the feature extraction-based recognition is highly vulnerable to the scalability adjustment, whereas the deep learning-based recognition shows robust performance regardless of the viewpoint.
In the following, therefore, the detailed performance analysis with respect to each video scalability dimension focuses on clusters 1 and 4.

\subsection{Quantization parameter}

When the spatial and temporal scalability options remain the same as in the original videos (i.e., 22 fps, 640$\times$480 pixels) and only the QP value changes, both the K-SVM and deep learning models maintain the recognition performance with tenacity.
For example, in cluster 1, the accuracies for the K-SVM model at $QP=0$, 36, 41, and 46 are 93.07, 90.15, 87.23, and 74.45\%, respectively, and those for the deep learning model are 87.23, 87.96, 83.21, and 79.20\%, respectively.
However, such endurance of the K-SVM model is beaten at $QP=51$; the accuracy is dropped to 28.83\%, while that for the deep learning model remains as 79.20\%.
The same tendency is observed for other spatiotemporal scalability configurations in both clusters 1 and 4.
This clearly demonstrates that the deep learning model is more robust to the visual degradation due to compression than the feature extraction-based model.

\begin{figure}[t]
	\centering
	\subfigure[]{
		\includegraphics[width=0.97in]{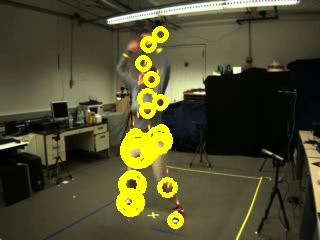}
		\label{fig:stip_per_qp_qp0}
	}
	\subfigure[]{
		\includegraphics[width=0.97in]{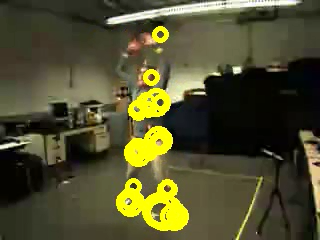}
		\label{fig:stip_per_qp_qp41}
	}
	\subfigure[]{
		\includegraphics[width=0.97in]{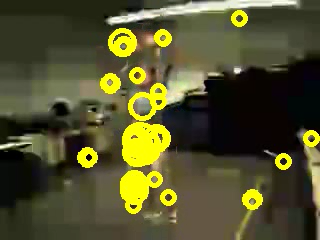}
		\label{fig:stip_per_qp_qp51}
	}
	\caption{Detected STIPs (represented by yellow circles) for an example video set for \textit{jumping jacks}. (a) $QP=0$ (b) $QP=41$ (c) $QP=51$}
	\label{fig:stip_per_qp}
\end{figure}

To discover the factors of performance changes, we investigate the features and output values generated from the feature extraction-based and deep learning models.
Figure~\ref{fig:stip_per_qp} shows STIPs detected for an example video set in cluster 1 for different QP values.
We can observe two problems that can cause lowered recognition performance for high QP values.
First, the detected feature points in a low quality video frame are at locations different from those in the original video frame; some feature points are lost due to the blurring artifacts and some are newly introduced due to the blocking artifacts.
For example, some STIP points in Figure~\ref{fig:stip_per_qp_qp51} are located outside the subject, which are irrelevant to the activity.
These points directly affect the histogram of codewords for a given video, which results in failure of the activity recognition.
Some pre-processing or image enhancement techniques could be used to alleviate this problem.
In addition, even when some feature points extracted from the videos having different QP values are in similar locations, the visual degradation directly affects to the values of the features.

\begin{figure*}[t]
	\centering
	\subfigure[]{
		\includegraphics[width=6.8in]{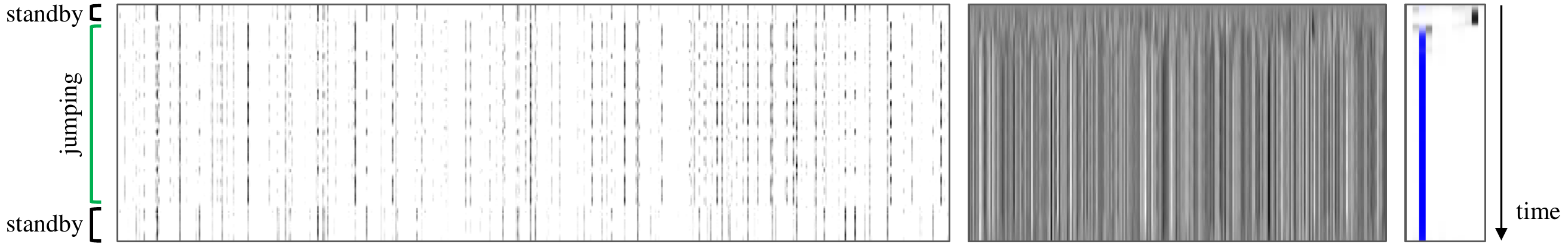}
		\label{fig:activation_per_qp_qp0}
	}
	\subfigure[]{
		\includegraphics[width=6.8in]{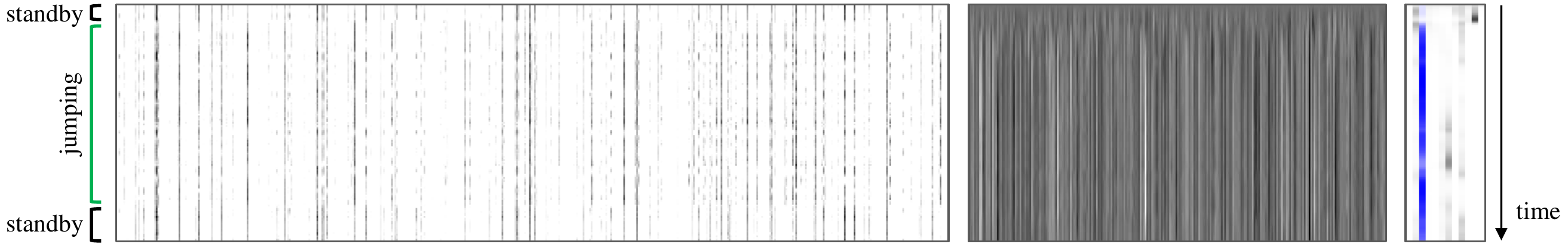}
		\label{fig:activation_per_qp_qp41}
	}
	\subfigure[]{
		\includegraphics[width=6.8in]{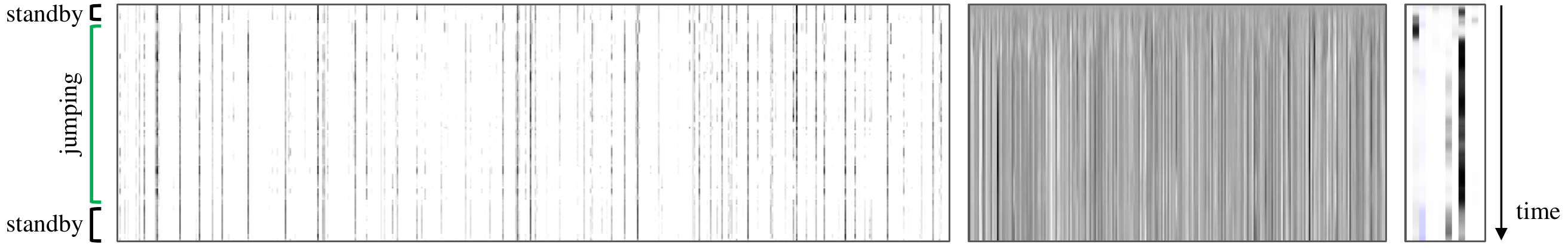}
		\label{fig:activation_per_qp_qp51}
	}
	\caption{Activations in the deep learning model for an example video set. (a) $QP=0$, (b) $QP=41$, and (c) $QP=51$. The figures on the left side show the activation values of the 512 neurons at \textit{full4}, those in the middle show the cell states of the 256 neurons at \textit{rnn7}, and those on the right side show the probability of each activity obtained from the 12 output neurons at \textit{full8}. A darker color means a higher value. In each figure, the activation of each neuron over time is represented as a column. In the figures on the right side, the output neuron corresponding to the ground-truth activity (i.e., \textit{jumping jacks}) is marked with blue colors.}
	\label{fig:activation_per_qp}
\end{figure*}

Figure~\ref{fig:activation_per_qp} shows activations of the deep learning model for an example video set in cluster 1 for different QP values.
It illustrates the output values of \textit{full4}, the cell states of \textit{rnn7}, and the final probabilities of the activity classes obtained from the softmax layer.
The videos represent the activity \textit{jumping jacks}.
In the videos, the subject is on standby for a moment, performs jumping jacks, and goes back to a standby at the end.
This behavior appears clearly on the left-side panel of Figure~\ref{fig:activation_per_qp_qp0}.
In the figure, the activation patterns at the beginning and end are very similar.
It is observed that during the jumping activity, two types of patterns alternate with each other as time goes, which appears as vertical dashed lines.
While the convolutional layers analyze these patterns in the spatial domain, the recurrent layers receive values from \textit{full4}, analyze in the temporal domain, and make decision with high consistency over time, as shown in the middle and right-side panels.

When the visual quality is degraded by increasing the QP value, the aforementioned patterns become unclear.
Altering the QP value from 0 to 41 makes the convolutional layers hard to detect alternating patterns; in the left-side panel of Figure~\ref{fig:activation_per_qp_qp41}, many dashed lines appearing in Figure~\ref{fig:activation_per_qp_qp0} are changed to solid lines.
Nevertheless, the recurrent layers capture the remaining alternating patterns and successfully recognize the activity correctly.
This shows that the cooperative operation of the convolutional and recurrent layers ensures the high robustness of the deep learning model against the visual quality degradation due to compression.
When the QP value becomes 51 (Figure~\ref{fig:activation_per_qp_qp51}), however, more spatial features are not captured, and the recurrent layers fail to classify the activity correctly.

\subsection{Spatial resolution}

\begin{figure*}[t]
	\centering
	\subfigure[]{
		\includegraphics[width=3.36in]{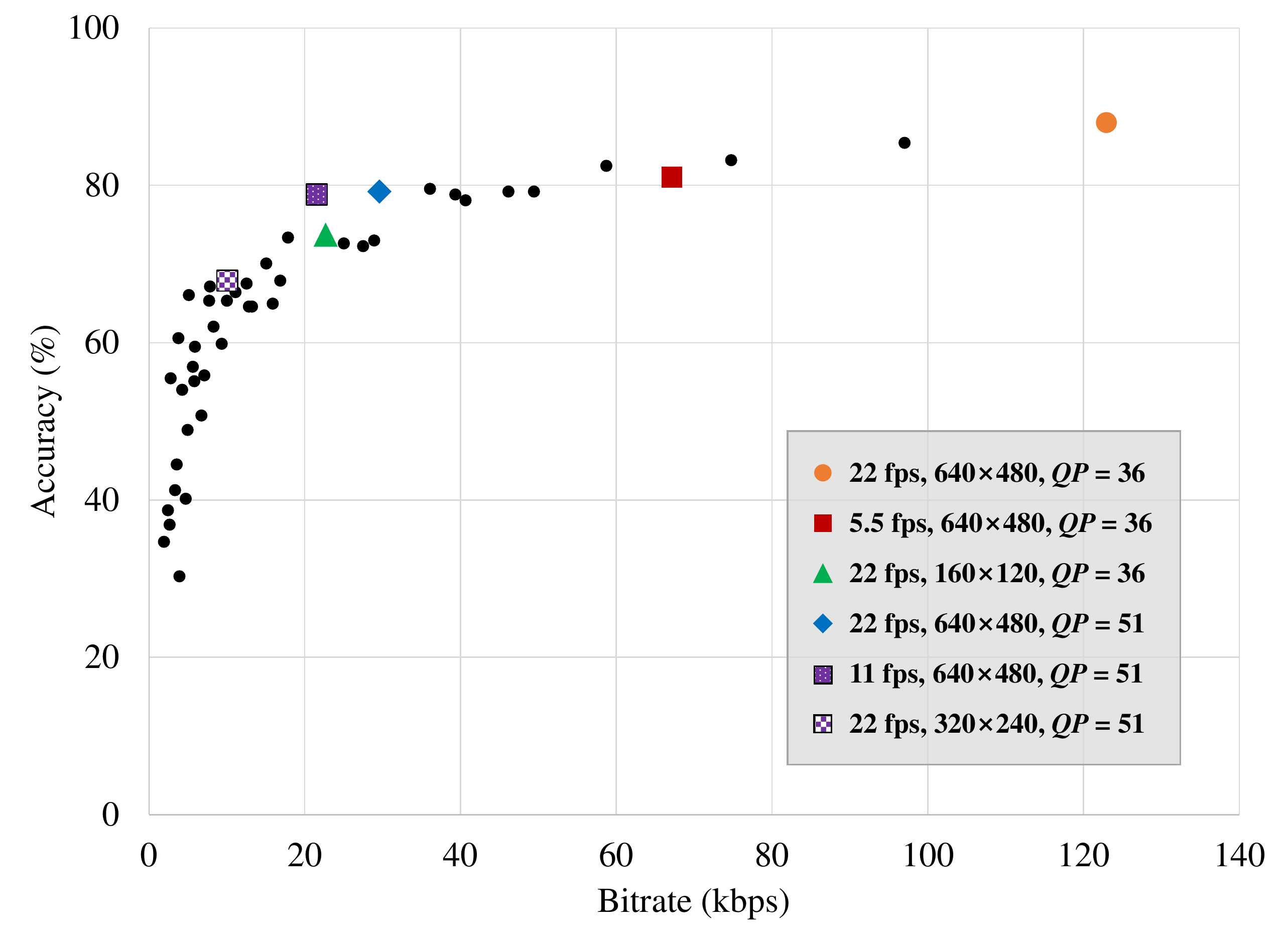}
	}
	\subfigure[]{
		\includegraphics[width=3.36in]{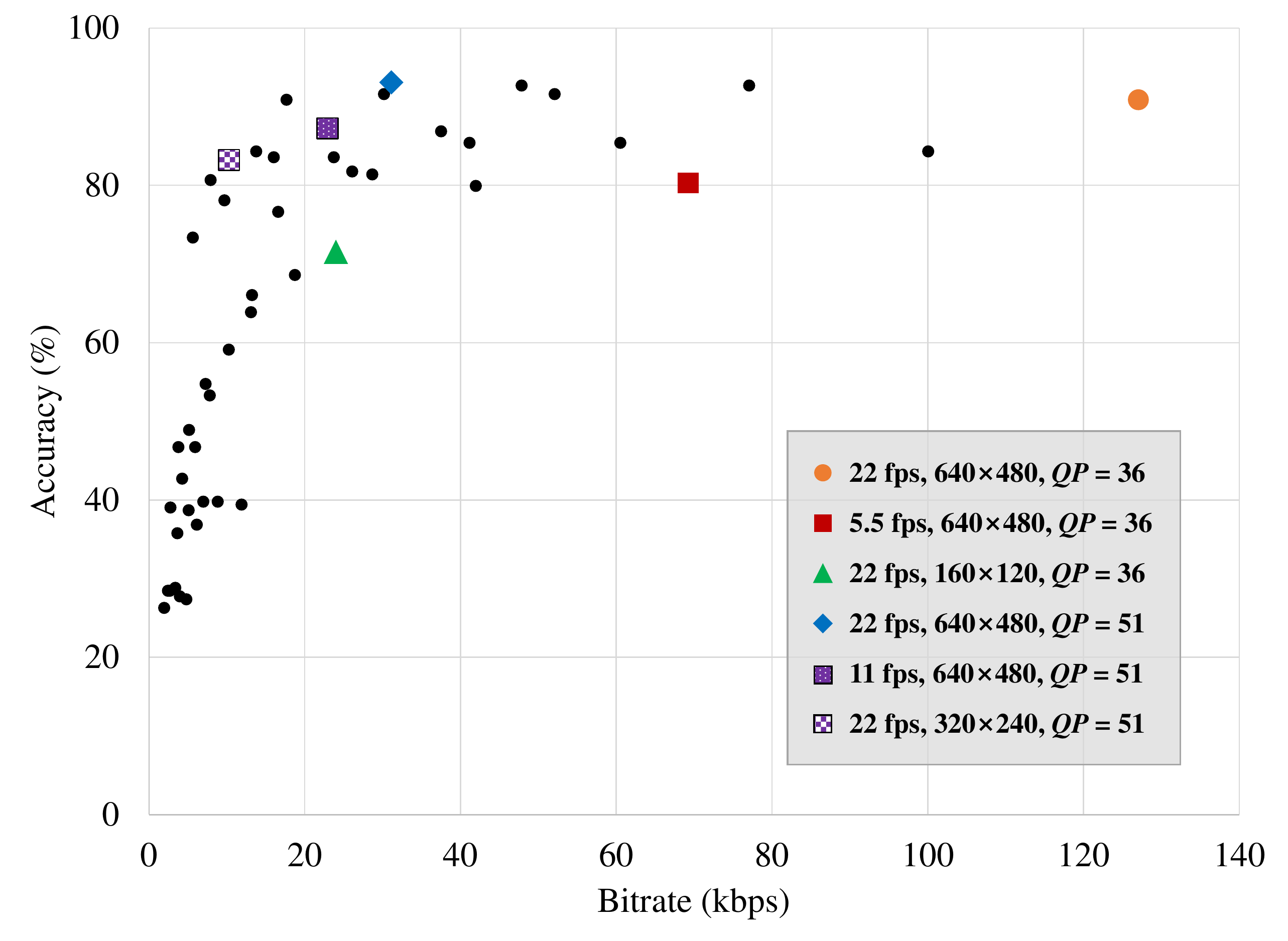}
	}
	\subfigure[]{
		\includegraphics[width=3.36in]{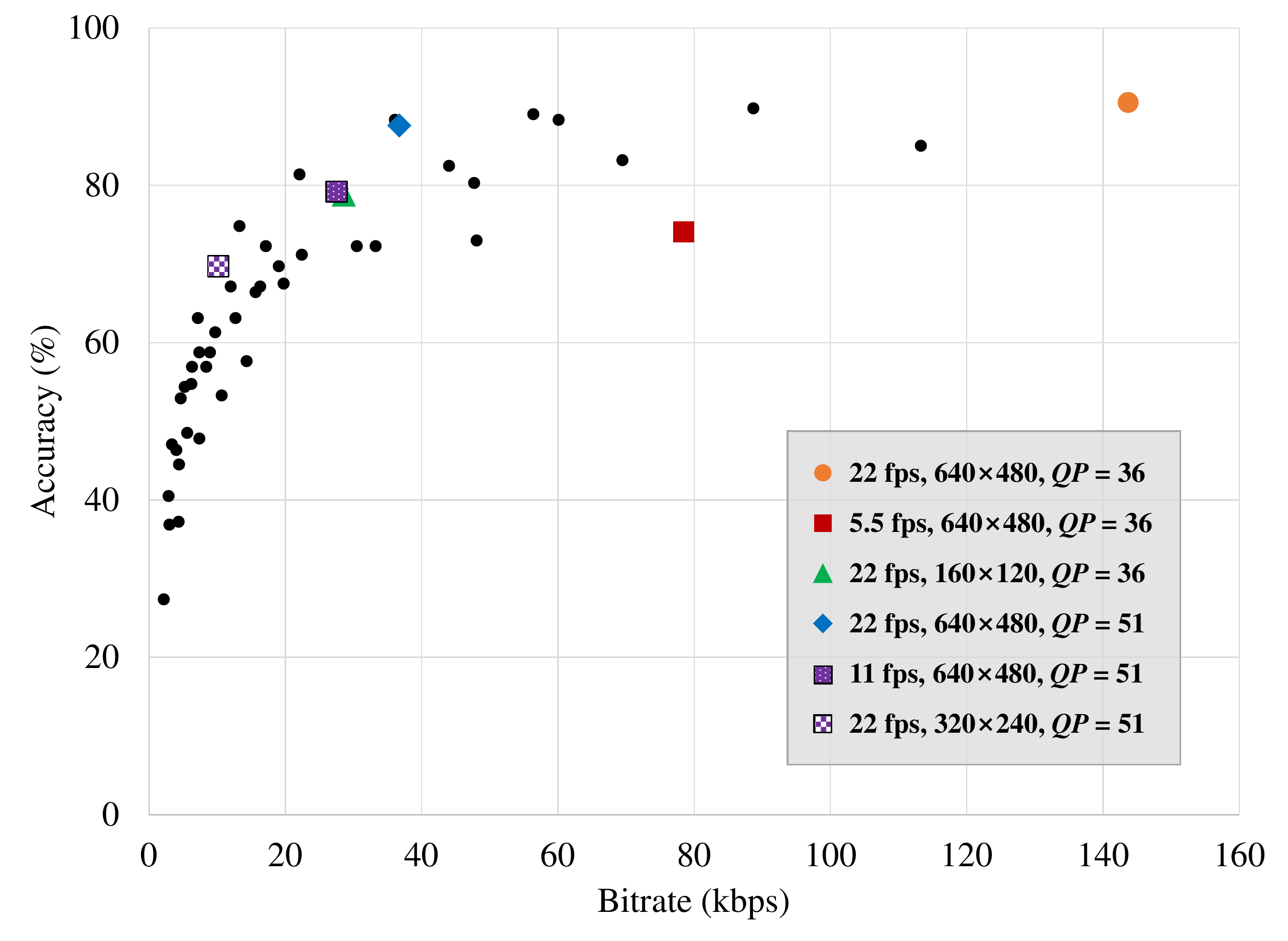}
	}
	\subfigure[]{
		\includegraphics[width=3.36in]{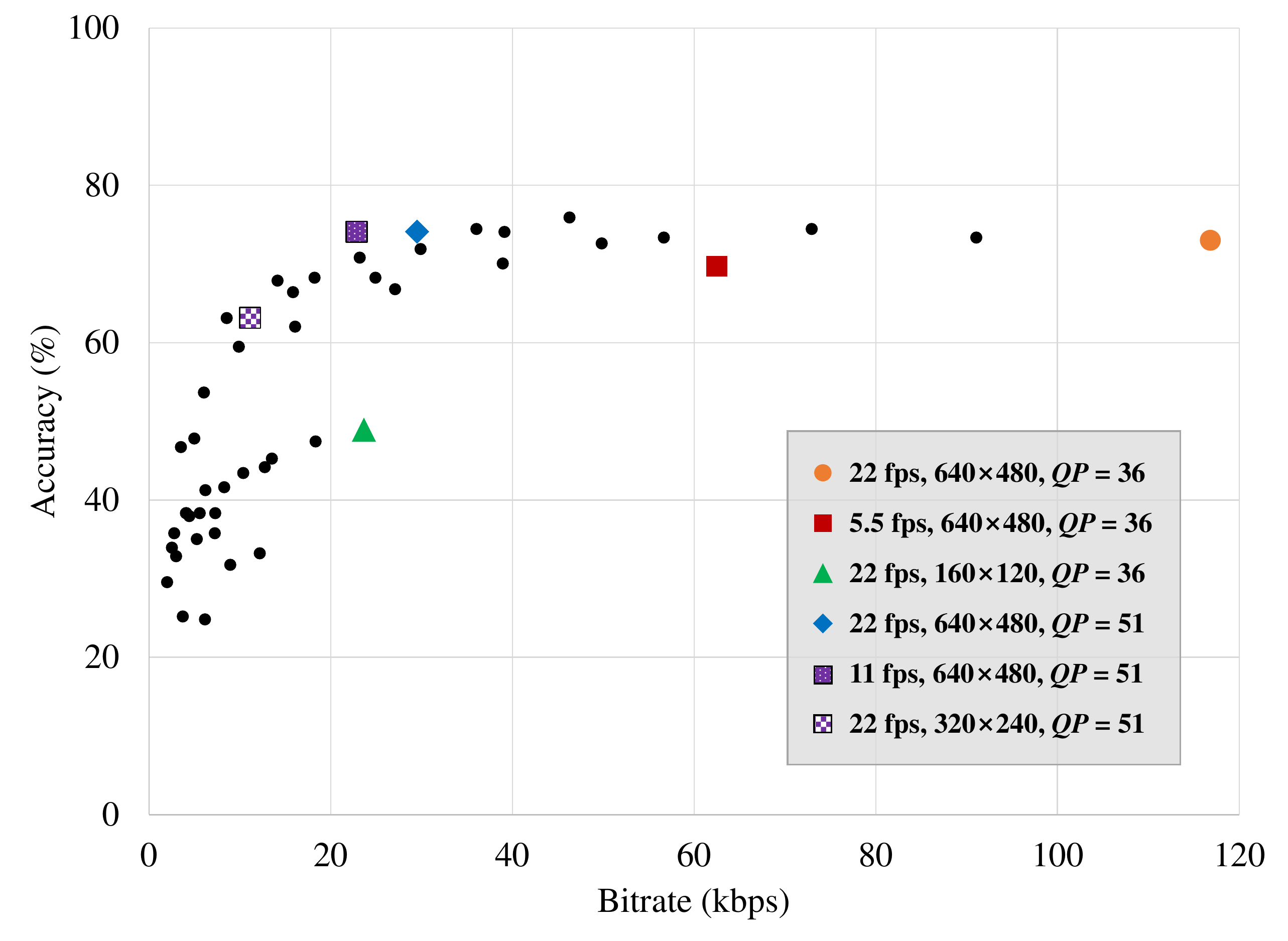}
	}
	\caption{Scatter plots between the average bitrates of the scalability combinations and the accuracies of the deep learning model. (a) Cluster 1 (b) Cluster 2 (c) Cluster 3 (d) Cluster 4}
	\label{fig:accuracy_vs_bitrate_clusters}
\end{figure*}

Overall, the K-SVM model fails to preserve its performance for reducing the spatial resolution in all cases, while the deep learning model does not.
When the frame size is reduced, all accuracies of the K-SVM model are below 40\% except for cluster 1, and the deep learning model outperforms the K-SVM model in all cases.
It is due to the high dependency of feature descriptors used in the K-SVM model on spatial features.

When the QP value is set to zero, reducing the resolution to 320$\times$240 pixels does not degrade the accuracy for the deep learning model, as expected in Section~\ref{sec:resolution}.
Starting from a resolution of 160$\times$120 pixels, the accuracy tends to decrease because the interpolation introduces blurring artifacts.
For example, in the case of cluster 1 with 22 fps, the accuracies for 640$\times$480, 320$\times$240, 160$\times$120, and 80$\times$60 pixels are 87.23, 85.04, 83.58, and 75.91\%, respectively.
Although the slight degradation in performance exists, the deep learning model still performs well in all cases; all accuracies exceed 60\% for all clusters.

However, when the visual degradation is additionally included by increasing the QP value, the performance degradation becomes prominent as the spatial resolution is reduced.
For example, in cluster 1 with 22 fps and $QP=51$, the accuracy of the deep learning model significantly drops to 30.29\% for a resolution of 80$\times$60 pixels from 75.91\% for 640$\times$480 pixels.
In addition, the performance change due to the spatial resolution reduction is more noticeable for higher QP values.
On the other hand, the performance degradation due to the spatial resolution reduction does not depend much on the frame rate.
These observations confirm that adjusting both spatial resolution and QP highly deteriorates the performance of the activity recognition models, since they both significantly affect the spatial characteristics.

\subsection{Frame rate}

It is observed that reducing the frame rate from 22 to 11 fps with preserving the spatial and quality resolutions does not significantly change the accuracy of the two models for clusters 1 and 4.
However, further reduction of the frame rate to 5.5 fps eventually decreases the accuracy of the K-SVM model, while the deep learning model keeps similar performance. 
For example, in cluster 1, the accuracies at 22, 11, and 5.5 fps for the K-SVM model are 93.07, 85.77, and 30.66\%, respectively, and those for the deep learning model are 87.23, 84.69, and 81.39\%, respectively.
This shows that the deep learning model is more robust than the feature extraction-based model even for the temporal degradation, along with the spatial degradation.

Since the frame rate is related to temporal characteristics rather than spatial ones, its influence on the performance acts almost independently on the other scalability dimensions.
For example, in cluster 1 with a resolution of 80$\times$60 pixels, lowering the frame rate from 22 to 5.5 fps changes the accuracy from 75.91 to 70.80\%, where the amount of reduced accuracy is almost the same to that without the spatial degradation.
The same tendency is observed when QP, instead of the spatial resolution, is reduced; in cluster 1 with $QP=51$, the accuracies for the frame rates of 22 and 5.5 fps are 79.20 and 70.07\%, respectively.

\subsection{Three-dimensional scalability}

Up to now, we focused on examining the influence of each scalability dimension.
In the following, we investigate how different combinations of scalability options affect the recognition performance.
This is an important issue because there exist multiple choices of scalability combinations giving similar bitrates and one needs to determine which scalability dimension should be adjusted to satisfy a given bitrate constraint and degrade recognition performance as little as possible.

Figure~\ref{fig:accuracy_vs_bitrate_clusters} shows the scatter plots between average bitrates of the scalability combinations and the accuracies of the deep learning model for each camera cluster.
The ideal cases, i.e., the combinations with a QP value of zero, are excluded.
Hence, the scalability option having the maximum bitrate is when the frame rate is 22 fps, the spatial resolution is 640$\times$480 pixels, and the QP value is 36, which is marked with an orange circle in each plot.
Overall, the accuracy tends to decrease naturally with respect to reduced bitrate.
However, it is clear that some scalability combinations have similar bitrates but yield different accuracies.

In Sections 4.1 to 4.3, it was shown that adjusting each video scalability option does not significantly degrade the overall performance of the deep learning model.
However, the significance of the bitrate reduction due to the scalability adjustment differs largely for each scalability dimension.
In the figure, some of these cases are marked as green triangles for the spatial degradation, red rectangles for the temporal degradation, and blue diamonds for the visual quality degradation.
When all the clusters are considered, increasing QP appears the best choice, which diminishes the largest amount of data but keeps the recognition performance almost unchanged.
Changing the spatial resolution can also reduce the bitrate significantly, but the recognition performance is lowered more severely.
Reduction of the frame rate leads to the smallest bitrate reduction, which seems to be due to the efficiency of the temporal prediction scheme in encoding.

Manipulating multiple scalability dimensions is helpful to achieve both reduced bitrate and reasonably good recognition performance.
For example, changing the frame rate to 11 fps and the QP value to 51, which is marked as purple rectangles, reduces the bitrate further, but achieves higher accuracies than the case showing similar bitrates via frame size reduction (i.e., 22 fps, 160$\times$120 pixels, $QP=36$, marked as green triangles in the figure).

If a much lower bitrate is required, adjusting the spatial resolution instead of the temporal resolution seems useful.
For example, the combination of a resolution of 320$\times$240 pixels and a QP value of 51, which is marked as half-filled purple rectangles, produces videos having even further reduced bitrates (about 10 kbps) with sacrificing the recognition performance slightly.
Finally, too much degradation in all scalability dimensions for extremely reduced bitrates clearly leads to excessively lowered performance, which correspond to the points shown in the lower left corner of each plot.

\section{Conclusion}

In this paper, we evaluated the performance of the activity recognition models with respect to multiple video scalability dimensions.
A K-SVM built with the STIP descriptor was used as the feature extraction-based model, and a deep neural network that consists of convolutional and recurrent layers was used as the deep learning model.
We tested the models with videos having various combinations of the video scalability options.

Our results showed that the deep learning model is highly robust against the degradation of the videos, in comparison to the feature extraction-based model.
Even though we did not train the deep learning model with degraded versions of the videos, it learned the useful characteristics of the activities with remarkable robustness against various types of video quality degradation.
Therefore, the deep learning model is suitable for real-world applications, which demand good recognition performance in limited resources.

In addition, we further investigated desirable scalability combinations for the deep learning model, which significantly reduce the amount of video data but guarantee competent recognition performance.
When only a single scalability dimension is adjusted, the visual quality degradation by increasing QP performs the best for both the bitrate reduction and the performance preservation.
Further improvements can be achieved by controlling multiple scalability dimensions, e.g., applying visual degradation with slightly lowered spatial or temporal resolutions.

\vfill

\section{Acknowledgments}
This work was supported by Institute for Information \& communications Technology Promotion (IITP) grant funded by the Korea government (MSIT) (R7124-16-0004, Development of Intelligent Interaction Technology Based on Context Awareness and Human Intention Understanding).

\bibliographystyle{abbrv}
\bibliography{sigproc} 

\end{document}